\documentclass[a4paper,11pt]{article}
\pdfoutput=1 

\usepackage{jheppub} 

\usepackage[T1]{fontenc} 

\title{\boldmath Clifford Algebras, Multipartite Systems and Gauge Theory Gravity}

\author[a]{Marco A. S. Trindade\note{Corresponding author.}}
\author[b]{Eric Pinto}
\author[c]{Sergio Floquet}

\affiliation[a]{Colegiado de F\'{i}sica \\
Departamento de Ci\^{e}ncias Exatas e da Terra\\
Universidade do Estado da Bahia,\\Salvador, BA, Brazil}
\affiliation[b]{Instituto de F\'{i}sica \\ Universidade Federal da Bahia\\ Salvador, BA, Brazil}
\affiliation[c]{Colegiado de Engenharia Civil\\Universidade Federal do Vale do S\~{a}o Francisco\\ Juazeiro, BA, Brazil.}

\emailAdd{matrindade@uneb.br}
\emailAdd{ericpinto3@gmail.com}
\emailAdd{sergio.floquet@univasf.edu.br}

\abstract{In this paper we present a multipartite formulation of gauge theory gravity based on the formalism of space-time algebra for gravitation developed by Lasenby and Doran (Lasenby, A. N., Doran, C. J. L, and  Gull, S.F.: Gravity, gauge theories and geometric algebra. Phil. Trans. R. Soc. Lond. A,  \textbf{582}, 356:487 (1998)). We associate the gauge fields with description of fermionic and bosonic states using the generalized graded tensor product. Einstein's equations are deduced from the graded projections and an algebraic Hopf-like structure naturally emerges from formalism. A connection with the theory of the quantum information is performed through the minimal left ideals and entangled qubits are derived. In addition applications to black holes physics and standard model are outlined.}

\begin{document}
\maketitle
\flushbottom

\section{Introduction}

This work is dedicated to the memory of professor Waldyr Alves Rodrigues Jr., whose contributions in the field of mathematical physics were of great prominence, especially in the study of clifford algebras and their applications to physics \cite{Waldyr1}. \

The gauge theory gravity\cite{Lasenby} is a gauge theory of gravitation based on geometric algebra formalism. Unlike the general relativity, it is formulated in a flat background spacetime with two basic principles related position-gauge invariance and rotation-gauge invariance so that gauge fields determine that relations between physical quantities are independent of the positions and orientations of the matter fields \cite{Lasenby}. This formulation presents some advantages in relation to general relativity. Interesting connections with quantum theory are extensively explored through the ubiquitous character of Clifford algebras although the issue of quantum gravitation is not explicitly discussed. \

Motivated by this theory we propose an extension of these ideas in a multipartite context on which we believe to help in possible connections with the quantum gravity (\hspace{-1pt}\cite{Doran}, p. 175). In the gauge theory gravity, the relation with quantum mechanics is mediated through the fact that the Dirac equation can be derived in a gravitational context using the gauge fields $g^{\mu \nu}$ and $\Omega^{\mu}$. Our multipartite formulation of gauge theory allows us to establish a new relationship with quantum mechanics since fields capable of describing bosonic and fermionic states have been derived.  We consider this important because the proposed relationship is valid for multiparticle systems and it is independent of the Dirac equation.  Besides, in the development of this formalism new algebraic structures were derived so that a way consistent with quantum mechanics was delineated. Our formalism carries the tensor product structure present in standard quantum mechanics where composite systems are described by tensor products of Hilbert spaces. Here, following an analogous way, the tensor products were considered, but exploring primitive elements and gradations in order to keep consistency with the gauge theory gravity. \

  In spite of the numerous proposals, the quantization of gravitational field remains one of major challenges of theoretical physics. The most promising proposals such as string theory \cite{strings1} and loop quantum gravity \cite{loop1} lack verifiable experimental predictions  with the current technological apparatus. The verification of the existence of extra dimensions and elementary quantum granular structure of space time at the Planck scale are some examples. Recently, a new approach builds a bridge between quantum gravity and quantum information theory through the correspondence AdS / CFT (or gauge/gravity duality) \cite{Ran1, Ran2, Ran3}. In this context, Einstein equations are derived from entanglement constraints for small perturbations to AdS. In our proposal, we obtained the Einstein equations in a multipartite context following the prescription of gauge theory gravity and considering graded projection operators \cite{Hestenes1, Hestenes2}  involving tensor products. For example, given $\nu$ multivectors $\Gamma_{\mu_{1}}, \Gamma_{\mu_{2}},\cdots ,\Gamma_{\mu_{\nu}}$, we define $\langle \sum_{\mu_{1}\mu_{2}\cdots \mu_{n}} \Gamma_{\mu_{1}} \otimes \Gamma_{\mu_{2}}\cdots \otimes \Gamma_{\mu_{\nu}}\rangle _{r_{1}-r_{2}-\cdots -r_{s}}$ as a sum in which the terms to be considered are only those of degree $r_{1}-r_{2}-\cdots -r_{s}$ without taking into account the order in the tensorial product. \

   From another point of view, but still related to ideas of quantum information, an emerging gravity model \cite{Verlinde1, Verlinde2} has been proposed in which gravity is not a fundamental interaction; it is an emergent phenomenon, an entropic force arising from the statistical behavior of microscopic degrees of freedom in a holographic scenario. It is postulated that each region of space is associated with a tensor factor of the microscopic Hilbert space and the associated entanglement entropy satisfies an area law in according to the Bekenstein-Hawking formula \cite{Verlinde2}. In this perspective it is important to highlight the seminal proposal of Sakharov, called induced gravity - a model of free non-interacting fields where the magnitude of the gravitational interaction is dictated by the masses and equations of motion for the free particles \cite{Sk}.\

   In this work the Clifford algebras and associated Hopf algebras \cite{Majid1, Castro, Majid2, Kassel, Chari} constitute the underlying basic structures. An interesting approach to quantum gravity involving Hopf algebras was performed by Majid \cite{Majid1, Majid2}. This formulation based on Hopf algebras contains a symmetry between states and observables with ideas of non-commutative geometry \cite{Majid1}. Majid argues that  both quantum theory and non-Euclidean geometry are needed for a self-dual picture which can be guaranteed with a bicrossproduct Hopf algebras structure.  Compatibility conditions involving bicrossproduct structure result in second-order gravitational field equations and the solutions of these equations corresponds to algebra of observables for the quantization of a particle moving on a homogeneous spacetime as highlighted in \cite{Majid1}. Still in this reference toy models are proposed and it is emphasized the need of generalization of the notion of Hopf algebras in order to obtain more realistic models. A recent proposal of space-time quantization in which Clifford algebras and quantum groups (quasitriangular Hopf algebras) play a key role was given by \cite{Hsu}. There are several interesting papers that explore the relationship between Clifford and Hopf algebras \cite{Roldao1, Lopez, Bulacu, Fauser1, Fauser2, Abla}. Particularly, relations between Clifford-Hopf algebras and quantum deformations of the Poincar\'{e} algebra were derived by Rocha, Bernardini and Vaz \cite{Roldao1}. In our formulation the relationship between Clifford and Hopf algebras arises from the need to use the primitive element $\Gamma_{\mu} \widehat{\otimes}1 +1 \widehat{\otimes}\Gamma_{\mu} $ in order to guarantee the consistency of the commutation and anti-commutation relations.  Here, we present the Hopf algebra associated to Clifford algebra through the generators of algebra and generalized graded tensor product, which is essential for obtaining the compatible co-algebra structure. \

   This paper is organized as follows: in section  \ref{secao2} we present the gauge theory gravity\cite{Lasenby} with typical structures of Clifford and Hopf algebras in an multipartite gravity model resulting in Einstein-like equations. Fields associated to bosonic and fermionic particles are deduced. Section  \ref{secao3} contains a connection with qubits and a deduction of entangled states in this formalism. In section  \ref{secao4} the Dirac equation is derived in the black holes background with the Schwarzschild solution following the reference \cite{Lasenby2}. In section \ref{secao5} our formalism is applied to Clifford algebra $Cl(0,6)$ describing to the $SU(3)_{c} \times SU(2)_{L} \times U(1)_{R} \times U(1)_{B-R} $ local gauge symmetries based on reference \cite{Wei}. In the last section we present our conclusions and perspectives.  \

\section{General formulation and Einstein field equations}\label{secao2}

   In the gauge theory gravity, the basic structure is space-time algebra $Cl(1,3)$ \cite{Lasenby}:
   \begin{eqnarray}
   \gamma_{\mu}\gamma_{\nu}+\gamma_{\nu} \gamma_{\mu}=2 \eta_{\mu \nu},
   \end{eqnarray}
   where $\eta_{\mu \nu}$ is the Minkowski with signature $(+---)$. Two gauge fields are introduced \cite{Lasenby}. The first is a vector field $g^{\mu}(x)$ with transformation law:
   \begin{eqnarray}
   g^{\mu}(x)\mapsto g^{\mu}(x)'=Rg^{\mu}(x)\widetilde{R},
   \end{eqnarray}
   where $R$ is a constant rotor. The second is the bivector-valued field whose law of transformation is given by
   \begin{eqnarray}
   \Omega_{\mu} \mapsto \Omega_{\mu}'=R \Omega_{\mu}\widetilde{R}-2\partial_{\mu} R\widetilde{R}
   \end{eqnarray}
   The covariant derivative is defined by:
   \begin{eqnarray}
   D_{\mu}\psi=\partial_{\mu}\psi+\frac{1}{2}\Omega_{\mu} \psi
   \end{eqnarray}
   The field gravitational equations are obtained from field strength through relation:
   \begin{eqnarray}
   [D_{\mu},D_{\nu}]\psi=\frac{1}{2}R_{\mu \nu}\psi
   \end{eqnarray}
   where
   \begin{eqnarray}
   R_{\mu \nu}=\partial_{\mu}\Omega_{\nu}-\partial_{\nu}\Omega_{\mu} + \Omega_{\mu} \times \Omega_{\nu}
   \end{eqnarray}
    An additional field equation is given by:
    \begin{eqnarray}
    D_{\mu}g_{\nu}-D_{\nu}g_{\mu}=0
    \end{eqnarray}

   One of the basic assumptions is that a quantum description of multipartite systems requires the structure of a tensor product associated with algebras analogously to the usual formulation of tensor products of Hilbert spaces in standard quantum mechanics. In order to implement such a structure preserving commutators and anticommutators we need to take into account primitive elements and the graded tensor product \cite{Majid2}. The start point is the generalized field strength:

\begin{eqnarray}
[D_{\mu}  \widehat{\otimes} 1\cdots \widehat{\otimes} 1+ \ \cdots  \ +1 \widehat{\otimes} 1 \widehat{\otimes} \cdots  \widehat{\otimes} D_{\mu}, D_{\nu} \widehat{\otimes} 1 \widehat{\otimes}\cdots \widehat{\otimes} 1+\ \cdots \ +1 \widehat{\otimes} 1 \widehat{\otimes} \cdots \widehat{\otimes} D_{\nu} ] \Psi \nonumber \\
=\frac{1}{2}\Psi(R_{\mu \nu} \widehat{\otimes} 1 \widehat{\otimes} \cdots \widehat{\otimes} 1 +\ \cdots \ + 1 \widehat{\otimes} 1 \widehat{\otimes}\cdots \widehat{\otimes} R_{\mu \nu}),
\end{eqnarray}
where the $\widehat{\otimes}$ is the generalized graded tensor product is given by
\begin{eqnarray}
(a_{1}\widehat{\otimes}a_{2}\widehat{\otimes}a_{3} \hspace{-0.3cm} &\cdots & \hspace{-0.3cm} \widehat{\otimes}a_{n-1}\widehat{\otimes}a_{n})
(b_{1}\widehat{\otimes}b_{2}\widehat{\otimes}b_{3}\cdots \widehat{\otimes}b_{n-1}\widehat{\otimes}b_{n}) \nonumber \\
 &=&  (-1)^{deg(a_{2})deg(b_{1})deg(a_{3})deg(b_{2})\cdots deg(a_{n})deg(b_{n-1})} \nonumber \\
 & & (a_{1}b_{1} \otimes a_{2}b_{2} \otimes\cdots \otimes a_{n}b_{n}). \label{pg}
\end{eqnarray}

In compact form
\begin{eqnarray}
[\widehat{D_{\mu}^{(n)}},\widehat{D_{\nu}^{(n)}}]  \Psi =\frac{1}{2} \widehat{R_{\mu \nu}^{(n)}} \Psi
\end{eqnarray}
with the generalized covariant derivative
\begin{eqnarray}
\widehat{D_{\mu}^{(n)}}&=&D_{\mu}\widehat{\otimes} 1 \widehat{\otimes} \cdots \widehat{\otimes} 1+ \ \cdots  \ +1 \widehat{\otimes} 1 \widehat{\otimes} \cdots  \widehat{\otimes} D_{\mu} \nonumber \\
&=&(\partial_{\mu}+\Omega_{\mu}\times)\widehat{\otimes} 1\cdots  \widehat{\otimes} 1+ \ \cdots  \ +1 \widehat{\otimes} 1 \widehat{\otimes}\cdots \widehat{\otimes}(\partial_{\mu}+\Omega_{\mu}\times) \nonumber
\end{eqnarray}
and
\begin{eqnarray}
\widehat{R_{\mu \nu}^{(n)}}&=&(\partial_{\mu}\Omega_{\nu}-\partial_{\nu}\Omega_{\mu}  + \Omega_{\mu} \times \Omega_{\nu})\widehat{\otimes} 1 \widehat{\otimes}\cdots \widehat{\otimes} 1 \nonumber \\
& &  + \ \cdots  \ +1 \widehat{\otimes} 1 \widehat{\otimes}\cdots \widehat{\otimes} (\partial_{\mu}\Omega_{\nu}-\partial_{\nu}\Omega_{\mu}  + \Omega_{\mu} \times \Omega_{\nu}) \nonumber
\end{eqnarray}
where there are $n$ terms. The Ricci tensor is given by

\begin{equation}
\widehat{R_{\nu}^{(n)}}=\langle \widehat{g^{\mu}_{(n)}} \widehat{R_{\mu \nu}^{(n)}} \rangle_{v-s}
\end{equation}
where $v-s$ is vector-scalar projection. Elements that have only these factors should be considered. Consequently
\begin{eqnarray}
\widehat{R_{\nu}^{(n)}}&=&\langle (g^{\mu} \widehat{\otimes} 1 \widehat{\otimes}\cdots \widehat{\otimes} 1+\ \cdots \ +1 \widehat{\otimes} 1 \cdots \widehat{\otimes} g^{\mu})(R_{\mu \nu} \widehat{\otimes} 1 \widehat{\otimes}\cdots \widehat{\otimes}1 \nonumber \\
& & +\ \cdots \ + 1 \widehat{\otimes} 1 \widehat{\otimes}\cdots  \widehat{\otimes}R_{\mu \nu})\rangle_{v-s} \nonumber \\
&=&\langle g^{\mu}R_{\mu \nu} \widehat{\otimes} 1 \widehat{\otimes} \cdots \widehat{\otimes} 1+g^{\mu} \widehat{\otimes} R_{\mu \nu} \widehat{\otimes}\cdots \widehat{\otimes} 1+\ \cdots \ +1 \widehat{\otimes} 1 \widehat{\otimes}\cdots \widehat{\otimes} g^{\mu}R_{\mu \nu} \rangle_{v-s} \nonumber \\
&=&g^{\mu}R_{\mu \nu} \widehat{\otimes} 1 \widehat{\otimes} \cdots \widehat{\otimes} 1+ \ \cdots \ + 1 \widehat{\otimes} 1 \widehat{\otimes}\cdots \widehat{\otimes} g^{\mu}R_{\mu \nu} \nonumber \\
&=&R_{\nu} \widehat{\otimes} 1 \widehat{\otimes}\cdots \widehat{\otimes} 1+\ \cdots \ +1 \widehat{\otimes} 1 \widehat{\otimes}\cdots \widehat{\otimes} R_{\nu}
\end{eqnarray}
It is verify that law of transformation for $\widehat{R_{\mu \nu}^{(n)}}$ is given by:
\begin{eqnarray}
\widehat{R_{\mu \nu}^{(n)}}'=\widetilde{R^{[n]}}\widehat{R_{\mu \nu}^{(n)}}R^{[n]},
\end{eqnarray}
where
\begin{eqnarray}
R^{[n]}=R \widehat{\otimes}R\widehat{\otimes}\cdots \widehat{\otimes}R
\end{eqnarray}
and
\begin{eqnarray}
\widetilde{R^{[n]}}=\widetilde{R}\widehat{\otimes}\widetilde{R}\widehat{\otimes}\cdots \widehat{\otimes}\widetilde{R}.
\end{eqnarray}
The Ricci scalar can obtained as:
\begin{eqnarray}
\widehat{R^{(n)}}=\langle \widehat{g^{\nu}_{(n)}}\widehat{R_{\nu}^{(n)}} \rangle_{s-s}
\end{eqnarray}
with
\begin{eqnarray}
\widehat{R^{(n)}}&=&\langle (g^{\nu} \widehat{\otimes} 1 \widehat{\otimes} \cdots \widehat{\otimes} 1+\ \cdots \ +1 \widehat{\otimes} 1 \widehat{\otimes} 1\cdots \widehat{\otimes} g^{\nu}) \nonumber \\
& &(g^{\nu} R_{\mu \nu} \widehat{\otimes} 1 \widehat{\otimes} \cdots \widehat{\otimes} 1+\ \cdots \ +1 \widehat{\otimes} 1 \widehat{\otimes} 1\cdots \widehat{\otimes} g^{\nu}R_{\mu \nu} )\rangle_{s-s}\nonumber \\
&=&g^{\nu} R_{\mu \nu} \widehat{\otimes} 1 \widehat{\otimes} \cdots \widehat{\otimes} 1+\ \cdots \ +1 \widehat{\otimes} 1 \widehat{\otimes} 1\cdots \widehat{\otimes} g^{\nu}R_{\mu \nu} \nonumber \\
&=&R \widehat{\otimes} 1 \widehat{\otimes} \cdots \widehat{\otimes} 1+\ \cdots \ +1 \widehat{\otimes} 1 \widehat{\otimes} \cdots \widehat{\otimes} R
\end{eqnarray}
Therefore the Einstein equations are given by
\begin{eqnarray}
\langle R_{\mu} \widehat{\otimes} 1 \widehat{\otimes}\cdots  \widehat{\otimes} 1  +\ \hspace{-0.3cm} &\cdots & \hspace{-0.3cm} \ +  1 \widehat{\otimes} 1 \cdots \widehat{\otimes} R_{\mu}
 -  \frac{1}{2}(R \widehat{\otimes}1 \widehat{\otimes}\cdots  \widehat{\otimes} 1+\cdots +1 \widehat{\otimes} 1 \widehat{\otimes}\cdots  \widehat{\otimes} R) \nonumber \\
  (g_{\mu} \widehat{\otimes} 1 \widehat{\otimes} \hspace{-0.3cm} &\cdots & \hspace{-0.3cm} \widehat{\otimes} 1 +\cdots + 1 \widehat{\otimes} 1 \widehat{\otimes} \cdots  \widehat{\otimes} g_{\mu})\rangle_{v-s} \nonumber \\
&=&8 \pi G \langle (T_{\mu} \widehat{\otimes} 1 \widehat{\otimes}\cdots \widehat{\otimes} 1+\ \cdots \ +1 \widehat{\otimes} 1 \widehat{\otimes}\cdots \widehat{\otimes} T_{\mu}) \rangle_{v-s} \nonumber
\end{eqnarray}
or
\begin{equation}
\widehat{R_{\mu}^{(n)}}-\frac{1}{2}\widehat{R^{(n)}}\widehat{g_{\mu}^{(n)}}=8 \pi \widehat{T_{\mu}^{(n)}}
\end{equation}
The Einstein equation appears as an emergent equation of a multipartite scenario from multivector fields. Explicitly, in this formulation we can derive fields associated to bosonic states related to bivector-valued field $\Omega_{\nu}$ through linear combinations as follows
\begin{eqnarray}
\sigma_{12\cdots i\cdots j\cdots n}^{(n)}&=&\sum_{\mu_{1},\mu_{2}}(\gamma_{\mu_{1}}^{1}\gamma_{\mu_{2}}^{1} \widehat{\otimes}1\widehat{\otimes} \cdots \widehat{\otimes} 1+\ \cdots \ +1 \widehat{\otimes} 1 \widehat{\otimes}\cdots \widehat{\otimes} \gamma_{\mu_{1}}^{1}\gamma_{\mu_{2}}^{1}) \nonumber \\
& &  (\gamma_{\mu_{1}}^{2}\gamma_{\mu_{2}}^{2} \widehat{\otimes}1 \widehat{\otimes}\cdots \widehat{\otimes} 1+\ \cdots \ +1 \widehat{\otimes} 1 \widehat{\otimes}\cdots \widehat{\otimes} \gamma_{\mu_{1}}^{2}\gamma_{\mu_{2}}^{2}) \nonumber \\
& & \ \cdots  \  \nonumber \\
& &  (\gamma_{\mu_{1}}^{i}\gamma_{\mu_{2}}^{i} \widehat{\otimes}1 \widehat{\otimes}\cdots \widehat{\otimes} 1+\ \cdots \ +1 \widehat{\otimes} 1 \widehat{\otimes}\cdots \widehat{\otimes} \gamma_{\mu_{1}}^{i}\gamma_{\mu_{2}}^{i}) \nonumber \\
& &  \cdots  \nonumber \\
& &   (\gamma_{\mu_{1}}^{j}\gamma_{\mu_{2}}^{j} \widehat{\otimes}1 \widehat{\otimes}\cdots \widehat{\otimes} 1+\ \cdots \ +1 \widehat{\otimes} 1 \widehat{\otimes}\cdots \widehat{\otimes} \gamma_{\mu_{1}}^{j}\gamma_{\mu_{2}}^{j}) \nonumber \\
& &\cdots  \nonumber \\
& &   (\gamma_{\mu_{1}}^{n}\gamma_{\mu_{2}}^{n} \widehat{\otimes}1\widehat{\otimes}\cdots \widehat{\otimes} 1+\ \cdots \ +1 \widehat{\otimes} 1 \widehat{\otimes}\cdots \widehat{\otimes} \gamma_{\mu_{1}}^{n}\gamma_{\mu_{2}}^{n}) \nonumber \\
\end{eqnarray}
with $\gamma_{\mu_{1}}^{k}, \gamma_{\mu_{2}}^{k} \neq \gamma_{\mu_{1}}^{l}, \gamma_{\mu_{2}}^{l} $ or $\gamma_{\mu_{1}}^{k} =  \gamma_{\mu_{1}}^{l}$ and $\gamma_{\mu_{2}}^{k} = \gamma_{\mu_{2}}^{l} $. Alternatively:
\begin{eqnarray}
\sigma_{12\cdots i\cdots j\cdots n}^{(n)}&=&\sum_{\mu_{1},\mu_{2}}(\gamma_{\mu_{1}}^{1} \widehat{\otimes}\gamma_{\mu_{1}}^{1}\widehat{\otimes} \cdots \widehat{\otimes} 1_{(2m+1)}+1_{(2m+1)} \widehat{\otimes} \gamma_{\mu_{1}}^{1}\widehat{\otimes}\cdots \widehat{\otimes} \gamma_{\mu_{1}}^{1}) \nonumber \\
& & (\gamma_{\mu_{2}}^{1} \widehat{\otimes}\gamma_{\mu_{2}}^{1}\widehat{\otimes} \cdots \widehat{\otimes} 1_{(2m+1)}+1_{(2m+1)} \widehat{\otimes} \gamma_{\mu_{2}}^{1}\widehat{\otimes}\cdots \widehat{\otimes} \gamma_{\mu_{2}}^{1}) \nonumber \\
& & (\gamma_{\mu_{1}}^{2} \widehat{\otimes}\gamma_{\mu_{1}}^{2} \widehat{\otimes}\cdots \widehat{\otimes} 1_{(2m+1)}+1_{(2m+1)} \widehat{\otimes} 1 \widehat{\otimes}\gamma_{\mu_{1}}^{2}\cdots \widehat{\otimes} \gamma_{\mu_{1}}^{2}) \nonumber \\
& & (\gamma_{\mu_{2}}^{2} \widehat{\otimes} \gamma_{\mu_{2}}^{2} \widehat{\otimes}\cdots \widehat{\otimes} 1_{(2m+1)}+1_{(2m+1)} \widehat{\otimes} \gamma_{\mu_{2}}^{2} \widehat{\otimes}\cdots \widehat{\otimes} \gamma_{\mu_{2}}^{2}) \nonumber \\
& &\cdots  \nonumber \\
& &   (\gamma_{\mu_{1}}^{i} \widehat{\otimes}\gamma_{\mu_{1}}^{i} \widehat{\otimes}\cdots \widehat{\otimes} 1_{(2m+1)}+1_{(2m+1)} \widehat{\otimes} \gamma_{\mu_{1}}^{i} \widehat{\otimes}\cdots \widehat{\otimes} \gamma_{\mu_{1}}^{i}) \nonumber \\
& &  (\gamma_{\mu_{2}}^{i} \widehat{\otimes}\gamma_{\mu_{2}}^{i} \widehat{\otimes}\cdots \widehat{\otimes} 1_{(2m+1)}+1_{(2m+1)} \widehat{\otimes} \gamma_{\mu_{2}}^{i} \widehat{\otimes}\cdots \widehat{\otimes} \gamma_{\mu_{2}}^{i}) \nonumber \\
& &\cdots  \nonumber \\
& & (\gamma_{\mu_{1}}^{j} \widehat{\otimes}\gamma_{\mu_{1}}^{j} \widehat{\otimes}\cdots \widehat{\otimes} 1_{(2m+1)}+1_{(2m+1)} \widehat{\otimes} \gamma_{\mu_{1}}^{j} \widehat{\otimes}\cdots \widehat{\otimes} \gamma_{\mu_{1}}^{j}) \nonumber \\
& &  (\gamma_{\mu_{2}}^{j} \widehat{\otimes}\gamma_{\mu_{2}}^{j} \widehat{\otimes}\cdots \widehat{\otimes} 1_{(2m+1)}+1_{(2m+1)} \widehat{\otimes} \gamma_{\mu_{2}}^{j} \widehat{\otimes}\cdots \widehat{\otimes} \gamma_{\mu_{2}}^{j}) \nonumber \\
& &\cdots  \nonumber \\
& &  (\gamma_{\mu_{1}}^{n} \widehat{\otimes}\gamma_{\mu_{1}}^{n} \widehat{\otimes}\cdots \widehat{\otimes} 1_{(2m+1)}+1_{(2m+1)} \widehat{\otimes} \gamma_{\mu_{1}}^{n} \widehat{\otimes}\cdots \widehat{\otimes} \gamma_{\mu_{1}}^{n}) \nonumber \\
& &  (\gamma_{\mu_{2}}^{n} \widehat{\otimes}\gamma_{\mu_{2}}^{n} \widehat{\otimes}\cdots \widehat{\otimes} 1_{(2m+1)}+1_{(2m+1)} \widehat{\otimes} \gamma_{\mu_{2}}^{n} \widehat{\otimes}\cdots \widehat{\otimes} \gamma_{\mu_{2}}^{n})
\end{eqnarray}
with $\gamma_{\mu_{1}}^{k}, \gamma_{\mu_{2}}^{k} \neq \gamma_{\mu_{1}}^{l}, \gamma_{\mu_{2}}^{l} $ or $\gamma_{\mu_{1}}^{k} =  \gamma_{\mu_{1}}^{l}$ and $\gamma_{\mu_{2}}^{k} = \gamma_{\mu_{2}}^{l} $. In order to ensure anticommutativity the number of factors must be even in the tensor product so that $n=2(2m+1)$, $m=0,1,2,\cdots $. We define the subscript $(2m+1)$ as $1_{(2m+1)}\equiv 1\widehat{\otimes} 1\widehat{\otimes}\cdots 1\widehat{\otimes}1$, $2m+1$ times, i.e., $2m+1$ factors. Analogously,
$a_{i} \widehat{\otimes}a_{i} \widehat{\otimes}\cdots \widehat{\otimes} 1_{(2m+1)}$ represent tensor product of $a_{i}$, $2m+1$ times, followed by the tensor product of $1$, $2m+1$ times,
while $1_{(2m+1)} \widehat{\otimes} a_{i} \widehat{\otimes}\cdots \widehat{\otimes} a_{i}$ represent tensor product of $1$, $2m+1$ times, followed by the tensor product of $a_{i}$, $2m+1$ times. 
It is easy verify that

\begin{eqnarray}
\sigma_{12\cdots j\cdots i\cdots n}^{(n)}&=&\sum_{\mu_{1},\mu_{2}}(\gamma_{\mu_{1}}^{1}\gamma_{\mu_{2}}^{1} \widehat{\otimes}1 \widehat{\otimes}\cdots \widehat{\otimes} 1+\ \cdots \ +1 \widehat{\otimes} 1 \widehat{\otimes}\cdots \widehat{\otimes} \gamma_{\mu_{1}}^{1}\gamma_{\mu_{2}}^{1}) \nonumber \\
& &  (\gamma_{\mu_{1}}^{2}\gamma_{\mu_{2}}^{2} \widehat{\otimes}1 \widehat{\otimes}\cdots \widehat{\otimes} 1+\ \cdots \ +1 \widehat{\otimes} 1 \widehat{\otimes}\cdots \widehat{\otimes} \gamma_{\mu_{1}}^{2}\gamma_{\mu_{2}}^{2}) \nonumber \\
& &\cdots  \nonumber \\
& &   (\gamma_{\mu_{1}}^{j}\gamma_{\mu_{2}}^{j} \widehat{\otimes} 1 \widehat{\otimes}\cdots \widehat{\otimes} 1+\ \cdots \ +1 \widehat{\otimes} 1 \widehat{\otimes}\cdots \widehat{\otimes} \gamma_{\mu_{1}}^{j}\gamma_{\mu_{2}}^{j}) \nonumber \\
& &\cdots  \nonumber \\
& &   (\gamma_{\mu_{1}}^{i}\gamma_{\mu_{2}}^{i} \widehat{\otimes} 1 \widehat{\otimes}\cdots \widehat{\otimes} 1+\ \cdots \ +1 \widehat{\otimes} 1 \widehat{\otimes}\cdots \widehat{\otimes} \gamma_{\mu_{1}}^{i}\gamma_{\mu_{2}}^{i}) \nonumber \\
& &\cdots  \nonumber \\
& &   (\gamma_{\mu_{1}}^{n}\gamma_{\mu_{2}}^{n} \widehat{\otimes} 1 \widehat{\otimes}\cdots \widehat{\otimes} 1+\ \cdots \ +1 \widehat{\otimes} 1 \widehat{\otimes}\cdots \widehat{\otimes} \gamma_{\mu_{1}}^{n}\gamma_{\mu_{2}}^{n}) \nonumber \\
&=&\sigma_{12\cdots i\cdots j\cdots n}^{(n)}
\end{eqnarray}

For the fields associated to fermionic states related to vector field $g^{\mu}(x)$ through linear combinations:

\begin{eqnarray}
\theta_{12\cdots i\cdots j\cdots n}^{(n)}&=&\sum_{\mu}(\gamma_{\mu}^{1} \widehat{\otimes} \gamma_{\mu}^{1} \widehat{\otimes}\cdots \widehat{\otimes} 1_{(2m+1)}+ 1_{(2m+1)} \widehat{\otimes}\gamma_{\mu}^{1} \cdots \widehat{\otimes} \gamma_{\mu}^{1}) \nonumber \\
& &  (\gamma_{\mu}^{2} \widehat{\otimes} \gamma_{\mu}^{2} \widehat{\otimes}\cdots \widehat{\otimes} 1_{(2m+1)}+ 1_{(2m+1)} \widehat{\otimes}\gamma_{\mu}^{2} \cdots \widehat{\otimes} \gamma_{\mu}^{2}) \nonumber \\
& &\cdots  \nonumber \\
& &   (\gamma_{\mu}^{i} \widehat{\otimes} \gamma_{\mu}^{i} \widehat{\otimes}\cdots \widehat{\otimes} 1_{(2m+1)}+ 1_{(2m+1)} \widehat{\otimes}\gamma_{\mu}^{i} \cdots \widehat{\otimes} \gamma_{\mu}^{i}) \nonumber \\
& &\cdots  \nonumber \\
& &   (\gamma_{\mu}^{j} \widehat{\otimes} \gamma_{\mu}^{j} \widehat{\otimes}\cdots \widehat{\otimes} 1_{(2m+1)}+ 1_{(2m+1)} \widehat{\otimes}\gamma_{\mu}^{j} \cdots \widehat{\otimes} \gamma_{\mu}^{j}) \nonumber \\
& &\cdots  \nonumber \\
& &  (\gamma_{\mu}^{n} \widehat{\otimes} \gamma_{\mu}^{n} \widehat{\otimes}\cdots \widehat{\otimes} 1_{(2m+1)}+ 1_{(2m+1)} \widehat{\otimes}\gamma_{\mu}^{n} \cdots \widehat{\otimes} \gamma_{\mu}^{n}) \nonumber \\
\end{eqnarray}

In order to ensure anti-commutativity the number of factors must be even in the tensor product. Analogously to bosonic case, it is easy verify that

\begin{eqnarray}
\theta_{12\cdots j\cdots i\cdots n}^{(n)}&=&\sum_{\mu}(\gamma_{\mu}^{1} \widehat{\otimes} \gamma_{\mu}^{1} \widehat{\otimes}\cdots \widehat{\otimes} 1_{(2m+1)}+ 1 _{(2m+1)}\widehat{\otimes}\gamma_{\mu}^{1} \cdots \widehat{\otimes} \gamma_{\mu}^{1}) \nonumber \\
& &   (\gamma_{\mu}^{2} \widehat{\otimes} \gamma_{\mu}^{2} \widehat{\otimes}\cdots \widehat{\otimes} 1_{(2m+1)}+ 1_{(2m+1)} \widehat{\otimes}\gamma_{\mu}^{2} \cdots \widehat{\otimes} \gamma_{\mu}^{2}) \nonumber \\
& &\cdots  \nonumber \\
& &   (\gamma_{\mu}^{j} \widehat{\otimes} \gamma_{\mu}^{j} \widehat{\otimes}\cdots \widehat{\otimes} 1_{(2m+1)}+ 1_{(2m+1)} \widehat{\otimes}\gamma_{\mu}^{j} \cdots \widehat{\otimes} \gamma_{\mu}^{j}) \nonumber \\
& &\cdots  \nonumber \\
& &  (\gamma_{\mu}^{i} \widehat{\otimes} \gamma_{\mu}^{i} \widehat{\otimes}\cdots \widehat{\otimes} 1_{(2m+1)}+ 1_{(2m+1)} \widehat{\otimes}\gamma_{\mu}^{i} \cdots \widehat{\otimes} \gamma_{\mu}^{i}) \nonumber \\
& &\cdots  \nonumber \\
& &   (\gamma_{\mu}^{n} \widehat{\otimes} \gamma_{\mu}^{n} \widehat{\otimes}\cdots \widehat{\otimes} 1_{(2m+1)}+ 1_{(2m+1)} \widehat{\otimes}\gamma_{\mu}^{n} \cdots \widehat{\otimes} \gamma_{\mu}^{n}) \nonumber \\
&=&-\theta_{12\cdots i\cdots j\cdots n}^{(n)}
\end{eqnarray}

The second field equation is given by:
\begin{eqnarray}
\hspace{-0.8cm} \langle(D_{\mu} \widehat{\otimes} 1 \widehat{\otimes}\cdots \widehat{\otimes}1+\cdots +1 \widehat{\otimes} 1 \widehat{\otimes}\cdots D_{\mu})
(g_{\nu} \widehat{\otimes} 1 \widehat{\otimes}\cdots \widehat{\otimes}1+\cdots +1 \widehat{\otimes} 1 \widehat{\otimes}\cdots g_{\nu})\qquad   \nonumber\\
\hspace{-0.8cm} -(D_{\nu} \widehat{\otimes} 1 \widehat{\otimes}\cdots \widehat{\otimes}1+\cdots +1 \widehat{\otimes} 1 \widehat{\otimes}\cdots D_{\nu})
(g_{\mu} \widehat{\otimes} 1 \widehat{\otimes}\cdots \widehat{\otimes}1+\cdots +1 \widehat{\otimes} 1 \widehat{\otimes}\cdots g_{\mu})\rangle_{v-s}  =0 \nonumber
\end{eqnarray}
where $\langle \ \rangle_{v-s}$ is the vector-scalar projection. In a compact form:
\begin{eqnarray}
\langle D_{\mu}^{(n)}g_{\nu}^{(n)}-D_{\nu}^{(n)}g_{\mu}^{(n)}\rangle_{v-s}=0
\end{eqnarray}

We have  an emergent Hopf algebra $(H,+,\cdot, \eta, \Delta, \epsilon, S,k)$ over $k$ \cite{Majid2, Kassel, Bulacu} of this formalism; where $\eta$ is unit map, $\Delta$ is the coproduct, $\epsilon$ is the counit and $S$ is antipode. The coproduct is defined by
\begin{eqnarray}
\Delta ( \gamma_{i}) = \frac{1}{\sqrt{2}}(\gamma_{i} \widehat\otimes 1  + 1 \widehat\otimes \gamma_{i})
\end{eqnarray}
and
\begin{eqnarray}
\Delta (1)= 1 \widehat{\otimes} 1
\end{eqnarray}
so that
\begin{eqnarray}
 \Delta (\{\gamma_{i}, \gamma_{j}\})&= &\Delta ( \gamma_{i})\Delta ( \gamma_{j})  + \Delta ( \gamma_{j})\Delta ( \gamma_{i}) \nonumber \\
&=& \Delta (2 \eta _{ij}1)=  2 \eta _{ij} (1 \widehat{\otimes} 1)
\end{eqnarray}

 We also have the definitions:
\begin{eqnarray}
\epsilon(\gamma_{i})&=&0, \ \ \ \ \ \ \   \epsilon(1)=1; \nonumber \\
S(\gamma_{i})&=&-\gamma_{i}, \ \ \  S(1)=1; \nonumber \\
\eta(1)&=&1, \ \ \ \ \ \ \    \eta(0)=0
\end{eqnarray}

The following axioms are valid:

1) Coalgebra:
\begin{eqnarray}
(\Delta \otimes id)\circ \Delta &=& (id \otimes \Delta) \circ \Delta \nonumber \\
(\epsilon \otimes id) \circ \Delta (c)&=& c = (id \otimes \epsilon) \circ \Delta (c),
\end{eqnarray}
with $c \in H$.
\

2) Bialgebra:
\begin{eqnarray}
\Delta(hg)&=&\Delta(h) \Delta(g), \ \ \Delta(1)=1 \widehat{\otimes} 1 \nonumber \\
\epsilon (gh)&=& \epsilon(g) \epsilon(h); \ \ \epsilon(1)=1
\end{eqnarray}
\

3) Hopf algebra (together with 1 and 2):

\begin{eqnarray}
 (S \otimes id)\circ \Delta = \cdot (id \otimes S) \circ \Delta = \eta \circ \epsilon
\end{eqnarray}

The elements of bosonic field can be obtained by coproduct $\Delta (\gamma_{i} \gamma_{j})$. Multipartite systems can be described making use of tensor product. This procedure can be generalized so that
\begin{eqnarray}
\Delta^{(n)} ( \gamma_{i}^{(2m+1)}) &=& \frac{1}{\sqrt{2}}(\gamma_{i} \widehat{\otimes} \gamma_{i} \widehat{\otimes}\cdots \widehat{\otimes} 1_{(2m+1)} + 1_{(2m+1)} \widehat{\otimes} \gamma_{i} \widehat{\otimes}\cdots \widehat{\otimes} \gamma_{i}) \nonumber \\
\Delta^{(n)} (1_{(2m+1)})&=& 1 \widehat{\otimes} 1 \widehat{\otimes}\cdots \widehat{\otimes}1\equiv 1_{2(2m+1)}
\end{eqnarray}
with an even number of factors in the tensor product, where $\gamma_{i}^{(2m+1)}\equiv\gamma_{i}\widehat{\otimes}\gamma_{i}\cdots \widehat{\otimes}\gamma_{i}$ $(2m+1)$ times . We have that
\begin{eqnarray}
\hspace{-0.5cm}  \Delta^{(n)} (\{\gamma_{i}^{(2m+1)}), \gamma_{j}^{(2m+1)})\})&=&\Delta^{(n)} ( \gamma_{i}^{(2m+1)}))\Delta^{(n)} ( \gamma_{j}^{(2m+1)})) \nonumber \\
\hspace{-0.5cm}  & & +\Delta^{(n)} ( \gamma_{j}^{(2m+1)}))\Delta^{(n)} ( \gamma_{i}^{(2m+1)})) \nonumber \\
 \hspace{-0.5cm} &=& \frac{1}{2}(\gamma_{i} \widehat{\otimes} \gamma_{i}\widehat{\otimes}\cdots \widehat{\otimes} 1_{(2m+1)}+ 1_{(2m+1)}\widehat{\otimes} \gamma_{i} \widehat{\otimes}\cdots \widehat{\otimes} \gamma_{i}) \nonumber \\
 & & (\gamma_{j} \widehat{\otimes} \gamma_{j}\widehat{\otimes}\cdots \widehat{\otimes} 1+ 1 \widehat{\otimes} \gamma_{j} \widehat{\otimes}\cdots \widehat{\otimes} \gamma_{j}) \nonumber \\
 & & + \frac{1}{2}(\gamma_{j} \widehat{\otimes} \gamma_{j}\widehat{\otimes}\cdots \widehat{\otimes} 1_{(2m+1)}+ 1_{(2m+1)} \widehat{\otimes} \gamma_{j} \widehat{\otimes}\cdots \widehat{\otimes} \gamma_{j}) \nonumber \\
 & &  (\gamma_{i} \widehat{\otimes} \gamma_{i}\widehat{\otimes}\cdots \widehat{\otimes} 1_{(2m+1)}+ 1_{(2m+1)} \widehat{\otimes} \gamma_{i} \widehat{\otimes}\cdots \widehat{\otimes} \gamma_{i}) \nonumber \\
&=& \Delta^{(n)} (2 \eta _{ij}1_{(2m+1)} ) \nonumber \\
&=&2 \eta _{ij} (1 \widehat{\otimes} 1 \widehat{\otimes}\cdots \widehat{\otimes} 1)\nonumber \\
&=&1_{2(2m+1)}
\end{eqnarray}

It is important to point out that this realization was possible due to the definition of the generalized graded tensor product (\ref{pg}).

\section{Qubits} \label{secao3}

In order to make a connection with quantum information, we consider the
bipartite case with the algebra called  $Cl^{(2)}(1,3)$:
\begin{eqnarray}
\{\Gamma_{\mu}, \Gamma_{\nu}\}=\{ \gamma_{\mu} \widehat{\otimes} 1 + 1\widehat{\otimes} \gamma_{\mu}, \gamma_{\nu} \widehat{\otimes} 1 + 1\widehat{\otimes} \gamma_{\nu}\}=2 \eta_{\mu \nu}(1 \widehat{\otimes} 1)
\end{eqnarray}
An primitive idempotent \cite{Rodrigues} associated with this algebra is given by
\begin{eqnarray}
\widehat{P}=\frac{1}{2}\left[1\widehat{\otimes} 1+\frac{1}{2}(\gamma_{3}\widehat{\otimes} 1+1 \widehat{\otimes} \gamma_{3} )(\gamma_{0}\widehat{\otimes}1+ 1\widehat{\otimes}\gamma_{0})\right].
\end{eqnarray}
Let be the isomorphism $Cl^{+}(1,3) \cong Cl(3,0)$:
\begin{eqnarray}
\sigma_{\mu}\in Cl(3,0) \ \leftrightarrow \ \gamma_{\mu}\gamma_{0} \in Cl^{+}(1,3)
\end{eqnarray}
In the $Cl(3,0)$ algebra, we have the primitive
\begin{eqnarray}
E=\frac{1}{2}(1+\sigma_{3})
\end{eqnarray}
Qubits can be identified as elements of minimal left ideals in these algebras \cite{Lasenby2, Baylis, Havel}:
\begin{eqnarray}
\gamma_{3}\gamma_{0}P &\leftrightarrow& \sigma_{3}E \ \leftrightarrow \ |0\rangle \nonumber \\
\gamma_{2}\gamma_{1}P &\leftrightarrow& \iota\sigma_{3}E \ \leftrightarrow \ i|0\rangle \nonumber \\
\gamma_{1}\gamma_{0}P &\leftrightarrow& \sigma_{1}E \ \leftrightarrow \ |1\rangle \nonumber \\
\gamma_{3}\gamma_{2}P &\leftrightarrow& \iota\sigma_{1}E \ \leftrightarrow \ i|1\rangle \nonumber
\end{eqnarray}
where $\iota=\sigma_{1}\sigma_{2}\sigma_{3}$ and $P=\frac{1}{2}(1+\gamma_{3}\gamma_{0})$. In the bipartite case, we have:
\begin{eqnarray}
T^{(1)}&=&(\gamma_{3}\widehat{\otimes} 1+1\widehat{\otimes} \gamma_{3})(\gamma_{0}\widehat{\otimes} 1+1\widehat{\otimes} \gamma_{0})(\gamma_{2}\widehat{\otimes} 1+1\widehat{\otimes} \gamma_{2})(\gamma_{1}\widehat{\otimes} 1+1\widehat{\otimes} \gamma_{1})\widehat{P} \leftrightarrow |00\rangle \nonumber \\
T^{(2)}&=&(\gamma_{1}\widehat{\otimes} 1+1\widehat{\otimes} \gamma_{1})(\gamma_{0}\widehat{\otimes} 1+1\widehat{\otimes} \gamma_{0})(\gamma_{3}\widehat{\otimes} 1+1\widehat{\otimes} \gamma_{3})(\gamma_{2}\widehat{\otimes} 1+1\widehat{\otimes} \gamma_{2})\widehat{P} \leftrightarrow |11\rangle \nonumber
\end{eqnarray}
so that
\begin{eqnarray}
\Psi=\frac{1}{\sqrt{2}} \left(T^{(1)}+T^{(2)}\right)\widehat{P}
\end{eqnarray}
corresponds to bipartite entangled state:
\begin{eqnarray}
|\Psi\rangle = \frac{(1+i)}{\sqrt{2}} \left(|00\rangle+|11\rangle\right)
\end{eqnarray}
Note that this is a possibility of building the fields so that commutativity is guaranteed. The bosonic field completely characterize the entangled qubits with the isomorphism $Cl^{(2)+}(1,3)\widehat{P} \simeq Cl(1,3)^{+}P \simeq Cl(3,0)E$. Explicitly, $\gamma_{j}\gamma_{0}\widehat{\otimes} 1+1\widehat{\otimes} \gamma_{j}\gamma_{0} \leftrightarrow \gamma_{j} \gamma_{0} \leftrightarrow \sigma_{j}$ [28]. Note that we perform the product by primitive idempotents so that we have minimal left ideals (algebraic spinors) \cite{Havel}. This procedure always can be generalized resulting in multipartite entangled states. \ \

\section{Black holes background} \label{secao4}
According to the prescription derived from our formalism, we deduced an equation of Dirac in a multipartite context in a way analogous to the reference \cite{Lasenby2}. So let us consider the Schwarzschild metric in Cartesian coordinates \cite{Lasenby}
\begin{eqnarray}
ds^{2}=\eta_{\mu \nu}dx^{\mu}dx^{\nu}-\frac{GM}{r}dt^{2}-\frac{2}{r}\left(\frac{2GM}{r}\right)^{1/2}b_{\mu}dtdx^{\mu},
\end{eqnarray}
where $b_{\mu}=(0,x,y,z)$ and $t$ is the proper time for an observer freely falling form rest at infinity. As highlighted in [26], the metric has interesting properties but it has been neglected. It is regular at $R=2M$, and it is singular at $r=0$. This reflects the fact that observers do not consider this surface as special. The surfaces $t=constant$ are flat so that the information about the spacetime curvature is therefore encoded in the off-diagonal component of the metric tensor \cite{Martel}. In this case, we have the gauge fields \cite{Lasenby, Lasenby2}
\begin{displaymath}
\left\{ \begin{array}{ll}
g_{0}=\gamma_{0}+\left(\frac{2GM}{r}\right)^{1/2}\frac{x^{i}}{r}\gamma_{i} \\
g_{i}=\gamma_{i},  \\
\Omega_{0}=\frac{G M}{r^2}    \sigma_{r} \\
\Omega_{i}=-\frac{1}{2r}\left(\frac{2GM}{r}\right)^{1/2}(2 \gamma_{i}\gamma_{0} -3\gamma_{i}\gamma_{0}\cdot \sigma_{r}\sigma_{r}),
\end{array} \right.
\end{displaymath}
where $ \sigma_{r}=\frac{1}{r}x^{i}\gamma_{i}\gamma_{0}$ and $i=1,2,3 $. In our formulation the Dirac equation is given by
\begin{eqnarray}
\hspace{-4.0cm}\langle (g^{\mu} \widehat{\otimes} 1 \widehat{\otimes}\cdots \widehat{\otimes} 1+\ \cdots \ +1 \widehat{\otimes} 1 \cdots \widehat{\otimes} g^{\mu})(D_{ \mu} \widehat{\otimes} 1 \widehat{\otimes}\cdots \widehat{\otimes} 1+\ \cdots \ + 1 \widehat{\otimes} 1 \widehat{\otimes}\cdots  \widehat{\otimes}D_{\mu }) \hspace{-4.5cm}& \nonumber \\
\hspace{-4.0cm}\cdot \Psi (I\sigma_{3}\widehat{\otimes} 1\cdots \widehat{\otimes} 1+\ \cdots \ +1\widehat{\otimes}I\sigma_{3})-m \Psi(\gamma_{0} \widehat{\otimes} 1\cdots \widehat{\otimes} 1+\ \cdots \ +1 \widehat{\otimes} 1\cdots \widehat{\otimes} \gamma_{0})\rangle_{v-v} \ = \ 0, \hspace{-4.5cm}& \nonumber \\
\end{eqnarray}
where $v-v$ is the vector-vector projection, i. e., only multivectors of grade-$1$ must be considered, analogously to the eq. (2.11). This equation can be rewritten as
\begin{eqnarray}
\sum_{\lambda}[(g^{\mu}D_{\mu}\Psi^{(1)}_{\lambda}I\sigma_{3} \hspace{-0.3cm} & - & \hspace{-0.3cm} m\Psi^{(1)}_{\lambda}\gamma_{0}) \widehat{\otimes} 1 \widehat{\otimes}\cdots \widehat{\otimes} 1  \nonumber \\
+ \ \cdots \ +1 \widehat{\otimes} 1 \hspace{-0.3cm} & \cdots  & \hspace{-0.3cm} \widehat{\otimes}(g^{\mu}D_{\mu}\Psi^{(n)}_{\lambda}I\sigma_{3}-m\Psi^{(n)}_{\lambda}\gamma_{0})] \ = \ 0 \nonumber
\end{eqnarray}
Therefore
\begin{eqnarray}
D \Psi &\equiv&\sum_{\lambda}\Biggl[\left(\bigtriangledown \Psi^{(1)}_{\lambda}-\gamma_{0}\left(\frac{2GM}{r}\right)^{1/2} \partial_{r}\Psi^{(1)}_{\lambda}-\frac{3}{4r} \left(\frac{2GM}{r}\right)^{1/2} \gamma_{0} \Psi^{(1)}_{\lambda} \right)I \sigma_{3} \nonumber \\
 & &- m \Psi^{(1)}_{\lambda} \gamma_{0}\Biggr] \widehat{\otimes} 1 \widehat{\otimes}\cdots \widehat{\otimes} 1+\ \cdots \ +1 \widehat{\otimes} 1 \cdots \widehat{\otimes}
 \Biggl[ \bigg(\bigtriangledown \Psi^{(n)}_{\lambda}   \nonumber \\
 &  & \left. -\gamma_{0}\left(\frac{2GM}{r}\right)^{1/2} \partial_{r}\Psi^{(n)}_{\lambda}-\frac{3}{4r} \left(\frac{2GM}{r}\right)^{1/2} \gamma_{0} \Psi^{(n)}_{\lambda} \right)I \sigma_{3}- m \Psi^{(n)}_{\lambda} \gamma_{0}\Biggr] \nonumber \\
 & =& 0
\end{eqnarray}

This is Dirac equation in the black holes background with our multipartite formalism. This equation describes the behavior of a system with many non-interacting fermions in the background black hole. The relevance of this issue was addressed in the reference \cite{Lasenby2},  where it was pointed out that a multiparticle approach is necessary to provide a proper link to the Hawking radiation. The existence of normalized bound state solutions associated to Dirac wave functions as well as non-hermiticity of the Hamiltonian associated to black hole singularity are carefully investigated issues in reference \cite{Lasenby2}. Here, we only intend to illustrate our formulation.

\section{Standard model} \label{secao5}
In this section we will explore the formalism previously developed in the context of the standard model. Our formulation is applied to approach based on the framework of Clifford algebra $Cl(0,6)$ to chiral symmetry breaking and
fermion mass hierarchies developed by Wei Lu \cite{Wei}. The $SU(3)_{c} \times SU(2)_{L} \times U(1)_{R} \times U(1)_{B-R} $ local gauge symmetries are characterized by this algebra. Thus, we consider the generators:
\begin{eqnarray}
\Gamma_{j}\Gamma_{k}+\Gamma_{j}\Gamma_{k}=-2 \delta_{jk}, \ \ \  j,k=1,2,\cdots 6
\end{eqnarray}
The generators of $Cl(1,3)$ subalgebra are given by \cite{Wei}
\begin{eqnarray}
\gamma_{0} &=& \Gamma_{1} \Gamma_{2} \Gamma_{3} \nonumber \\
\gamma_{1} &=& \Gamma_{4} \nonumber \\
\gamma_{2} &=& \Gamma_{5} \nonumber \\
\gamma_{3} &=& \Gamma_{6} \nonumber
\end{eqnarray}
The unit pseudoscalar is $i=\Gamma_{1}\Gamma_{2}\Gamma_{3}\Gamma_{4}\Gamma_{5}\Gamma_{6}=\gamma_{0}\gamma_{1}\gamma_{2}\gamma_{3}$. The $SU(3)_{c}$ strong interaction is given by $G_{\mu}^{(n)}=G_{\mu}^{k}T_{k}^{(n)}$, where the generators:
\begin{eqnarray}
T_{1}&=&\frac{1}{4}(\gamma_{1}\Gamma_{2}+\gamma_{2}\Gamma_{1})\widehat{\otimes} 1\widehat{\otimes} \cdots \widehat{\otimes} 1+\ \cdots \ +1\widehat{\otimes} 1\widehat{\otimes}\cdots \widehat{\otimes}\frac{1}{4}(\gamma_{1}\Gamma_{2}+\gamma_{2}\Gamma_{1}) \nonumber \\
T_{2}&=&\frac{1}{4}(\Gamma_{1}\Gamma_{2}+\gamma_{1}\gamma_{2})\widehat{\otimes} 1\widehat{\otimes}\cdots \widehat{\otimes} 1+\ \cdots \ +1\widehat{\otimes} 1\widehat{\otimes}\cdots \widehat{\otimes}\frac{1}{4}(\Gamma_{1}\Gamma_{2}+\gamma_{1}\gamma_{2}) \nonumber \\
T_{3}&=&\frac{1}{4}(\Gamma_{1}\gamma_{1}-\Gamma_{2}\gamma_{2})\widehat{\otimes} 1\widehat{\otimes}\cdots \widehat{\otimes} 1+\ \cdots \ +1\widehat{\otimes} 1\widehat{\otimes}\cdots \widehat{\otimes}\frac{1}{4}(\Gamma_{1}\gamma_{1}-\Gamma_{2}\gamma_{2}) \nonumber \\
T_{4}&=&\frac{1}{4}(\gamma_{1}\Gamma_{3}+\gamma_{3}\Gamma_{1})\widehat{\otimes} 1\widehat{\otimes}\cdots \widehat{\otimes} 1+\ \cdots \ +1\widehat{\otimes} 1\widehat{\otimes}\cdots \widehat{\otimes}\frac{1}{4}(\gamma_{1}\Gamma_{3}+\gamma_{3}\Gamma_{1}) \nonumber \\
T_{5}&=&\frac{1}{4}(\Gamma_{1}\Gamma_{3}+\gamma_{1}\gamma_{3})\widehat{\otimes} 1\widehat{\otimes}\cdots \widehat{\otimes} 1+\ \cdots \ +1\widehat{\otimes} 1\widehat{\otimes}\cdots \widehat{\otimes}\frac{1}{4}(\Gamma_{1}\Gamma_{3}+\gamma_{1}\gamma_{3}) \nonumber \\
T_{6}&=&\frac{1}{4}(\gamma_{2}\Gamma_{3}+\gamma_{3} \Gamma_{2})\widehat{\otimes} 1\widehat{\otimes}\cdots \widehat{\otimes} 1+\ \cdots \ +1\widehat{\otimes} 1\widehat{\otimes}\cdots \widehat{\otimes}\frac{1}{4}(\gamma_{2}\Gamma_{3}+\gamma_{3} \Gamma_{2}) \nonumber \\
T_{7}&=&\frac{1}{4}(\Gamma_{2}\Gamma_{3}+\gamma_{2}\gamma_{3})\widehat{\otimes} 1\widehat{\otimes}\cdots \widehat{\otimes} 1+\ \cdots \ +1\widehat{\otimes} 1\widehat{\otimes}\cdots \widehat{\otimes}\frac{1}{4}(\Gamma_{2}\Gamma_{3}+\gamma_{2}\gamma_{3}) \nonumber \\
T_{8}&=&\frac{1}{4\sqrt{3}}(\Gamma_{1}\gamma_{1}+\Gamma_{2}\gamma_{2}-2 \Gamma_{3}\gamma_{3})\widehat{\otimes} 1\widehat{\otimes}\cdots \widehat{\otimes}1 \nonumber \\
 & & +\ \cdots \ +1\widehat{\otimes} 1\widehat{\otimes}\cdots \widehat{\otimes}\frac{1}{4}(\Gamma_{1}\gamma_{1}+\Gamma_{2}\gamma_{2}-2 \Gamma_{3}\gamma_{3}).
\end{eqnarray}

 Alternatively,

\begin{eqnarray}
T_{1}&=&\frac{1}{4}[\gamma_{1}\widehat{\otimes} \gamma_{1}\widehat{\otimes}\cdots \widehat{\otimes}1_{(2m+1)}+ 1_{(2m+1)}\widehat{\otimes}\gamma_{1}\widehat{\otimes}\cdots \widehat{\otimes}\gamma_{1}] \nonumber \\
& & [\Gamma_{2}\widehat{\otimes} \Gamma_{2}\widehat{\otimes}\cdots \widehat{\otimes}1_{(2m+1)}+1_{(2m+1)}\widehat{\otimes}\Gamma_{2}\widehat{\otimes}\cdots \widehat{\otimes}\Gamma_{2}] \nonumber \\
& &+ \frac{1}{4}[\gamma_{2}\widehat{\otimes} \gamma_{2}\widehat{\otimes}\cdots \widehat{\otimes}1_{(2m+1)}+ 1_{(2m+1)}\widehat{\otimes} \gamma_{2}\widehat{\otimes}\cdots \widehat{\otimes}\gamma_{2}] \nonumber \\
& & [\Gamma_{1}\widehat{\otimes} \Gamma_{1}\widehat{\otimes}\cdots \widehat{\otimes}1_{(2m+1)}+ 1_{(2m+1)}\widehat{\otimes}\Gamma_{1}\widehat{\otimes}\cdots \widehat{\otimes}\Gamma_{1}]
\nonumber \\
T_{2}&=&\frac{1}{4}[\Gamma_{1}\widehat{\otimes} \Gamma_{1}\widehat{\otimes}\cdots \widehat{\otimes}1_{(2m+1)}+ 1_{(2m+1)}\widehat{\otimes}\Gamma_{1}\widehat{\otimes}\cdots \widehat{\otimes}\Gamma_{1}] \nonumber \\
& & [\Gamma_{2}\widehat{\otimes} \Gamma_{2}\widehat{\otimes}\cdots \widehat{\otimes}1_{(2m+1)}+1_{(2m+1)}\widehat{\otimes}\Gamma_{2}\widehat{\otimes}\cdots \widehat{\otimes}\Gamma_{2}] \nonumber \\
& &+\frac{1}{4}[\gamma_{1}\widehat{\otimes} \gamma_{1}\widehat{\otimes}\cdots \widehat{\otimes}1_{(2m+1)}+1_{(2m+1)}\widehat{\otimes}  \gamma_{1}\widehat{\otimes}\cdots \widehat{\otimes}\gamma_{1}] \nonumber \\
& & [\gamma_{2}\widehat{\otimes} \gamma_{2}\widehat{\otimes}\cdots \widehat{\otimes}1_{(2m+1)}+ 1_{(2m+1)}\widehat{\otimes}\gamma_{2}\widehat{\otimes}\cdots \widehat{\otimes}\gamma_{2}]
\nonumber
\end{eqnarray}
\begin{eqnarray}
T_{3}&=&\frac{1}{4}[\Gamma_{1}\widehat{\otimes} \Gamma_{1}\widehat{\otimes}\cdots \widehat{\otimes}1_{(2m+1)}+ 1_{(2m+1)}\widehat{\otimes}\Gamma_{1}\widehat{\otimes}\cdots \widehat{\otimes}\Gamma_{1}] \nonumber \\
& & [\gamma_{1}\widehat{\otimes} \gamma_{1}\widehat{\otimes}\cdots \widehat{\otimes}1_{(2m+1)}+1_{(2m+1)}\widehat{\otimes}\gamma_{1}\widehat{\otimes}\cdots \widehat{\otimes}\gamma_{1}] \nonumber \\
& & -\frac{1}{4}[\Gamma_{2}\widehat{\otimes} \Gamma_{2}\widehat{\otimes}\cdots \widehat{\otimes}1_{(2m+1)}+1_{(2m+1)}\widehat{\otimes} \Gamma_{2}\widehat{\otimes}\cdots \widehat{\otimes}\Gamma_{2}] \nonumber \\
& &  [\gamma_{2}\widehat{\otimes} \gamma_{2}\widehat{\otimes}\cdots \widehat{\otimes}1_{(2m+1)}+ 1_{(2m+1)}\widehat{\otimes}\gamma_{2}\widehat{\otimes}\cdots \widehat{\otimes}\gamma_{2}] \nonumber \\
T_{4}&=&\frac{1}{4}[\gamma_{1}\widehat{\otimes} \gamma_{1}\widehat{\otimes}\cdots \widehat{\otimes}1_{(2m+1)}+ 1_{(2m+1)}\widehat{\otimes}\gamma_{1}\widehat{\otimes}\cdots \widehat{\otimes}\gamma_{1}] \nonumber \\
& &  [\Gamma_{3}\widehat{\otimes} \Gamma_{3}\widehat{\otimes}\cdots \widehat{\otimes}1_{(2m+1)}+1_{(2m+1)}\widehat{\otimes}\Gamma_{3}\widehat{\otimes}\cdots \widehat{\otimes}\Gamma_{3}] \nonumber \\
& & +\frac{1}{4}[\gamma_{3}\widehat{\otimes} \gamma_{3}\widehat{\otimes}\cdots \widehat{\otimes}1_{(2m+1)}+ 1_{(2m+1)}\widehat{\otimes} \gamma_{3}\widehat{\otimes}\cdots \widehat{\otimes}\gamma_{3}] \nonumber \\
& & [\Gamma_{1}\widehat{\otimes} \Gamma_{1}\widehat{\otimes}\cdots \widehat{\otimes}1_{(2m+1)}+ 1_{(2m+1)}\widehat{\otimes}\Gamma_{1}\widehat{\otimes}\cdots \widehat{\otimes}\Gamma_{1}]
\nonumber \\
T_{5}&=&\frac{1}{4}[\Gamma_{1}\widehat{\otimes} \Gamma_{1}\widehat{\otimes}\cdots \widehat{\otimes}1_{(2m+1)}+ 1_{(2m+1)}\widehat{\otimes}\Gamma_{1}\widehat{\otimes}\cdots \widehat{\otimes}\Gamma_{1}] \nonumber \\
& &[\Gamma_{3}\widehat{\otimes} \Gamma_{3}\widehat{\otimes}\cdots \widehat{\otimes}1_{(2m+1)}+1_{(2m+1)}\widehat{\otimes}\Gamma_{3}\widehat{\otimes}\cdots \widehat{\otimes}\Gamma_{3}] \nonumber \\
& & +\frac{1}{4}[\gamma_{1}\widehat{\otimes} \gamma_{1}\widehat{\otimes}\cdots \widehat{\otimes}1_{(2m+1)}+ 1_{(2m+1)}\widehat{\otimes} \gamma_{1}\widehat{\otimes}\cdots \widehat{\otimes}\gamma_{1}] \nonumber \\
& & [\gamma_{3}\widehat{\otimes} \gamma_{3}\widehat{\otimes}\cdots \widehat{\otimes}1_{(2m+1)}+ 1_{(2m+1)}\widehat{\otimes}\gamma_{3}\widehat{\otimes}\cdots \widehat{\otimes}\gamma_{3}]
\nonumber \\
T_{6}&=&\frac{1}{4}[\gamma_{2}\widehat{\otimes} \gamma_{2}\widehat{\otimes}\cdots \widehat{\otimes}1_{(2m+1)}+ 1_{(2m+1)}\widehat{\otimes}\gamma_{2}\widehat{\otimes}\cdots \widehat{\otimes}\gamma_{2}] \nonumber \\
& & [\Gamma_{3}\widehat{\otimes} \Gamma_{3}\widehat{\otimes}\cdots \widehat{\otimes}1_{(2m+1)}+1_{(2m+1)}\widehat{\otimes}\Gamma_{3}\widehat{\otimes}\cdots \widehat{\otimes}\Gamma_{3}] \nonumber \\
& & +\frac{1}{4}[\gamma_{3}\widehat{\otimes} \gamma_{3}\widehat{\otimes}\cdots \widehat{\otimes}1_{(2m+1)}+ 1_{(2m+1)}\widehat{\otimes} \gamma_{3}\widehat{\otimes}\cdots \widehat{\otimes}\gamma_{3}] \nonumber \\
& & [\Gamma_{2}\widehat{\otimes} \Gamma_{2}\widehat{\otimes}\cdots \widehat{\otimes}1_{(2m+1)}+ 1_{(2m+1)}\widehat{\otimes}\Gamma_{2}\widehat{\otimes}\cdots \widehat{\otimes}\Gamma_{2}]
\nonumber \\
T_{7}&=&\frac{1}{4}[\Gamma_{2}\widehat{\otimes} \Gamma_{2}\widehat{\otimes}\cdots \widehat{\otimes}1_{(2m+1)}+ 1_{(2m+1)}\widehat{\otimes}\Gamma_{2}\widehat{\otimes}\cdots \widehat{\otimes}\Gamma_{2}] \nonumber \\
& & [\Gamma_{3}\widehat{\otimes} \Gamma_{3}\widehat{\otimes}\cdots \widehat{\otimes}1_{(2m+1)}+1_{(2m+1)}\widehat{\otimes}\Gamma_{3}\widehat{\otimes}\cdots \widehat{\otimes}\Gamma_{3}] \nonumber \\
& & +\frac{1}{4}[\gamma_{2}\widehat{\otimes} \gamma_{2}\widehat{\otimes}\cdots \widehat{\otimes}1_{(2m+1)}+ 1_{(2m+1)}\widehat{\otimes} \gamma_{2}\widehat{\otimes}\cdots \widehat{\otimes}\gamma_{2}] \nonumber \\
& & [\gamma_{3}\widehat{\otimes} \gamma_{3}\widehat{\otimes}\cdots \widehat{\otimes}1_{(2m+1)}+ 1_{(2m+1)}\widehat{\otimes}\gamma_{3}\widehat{\otimes}\cdots \widehat{\otimes}\gamma_{3}]
\nonumber \\
T_{8}&=&\frac{1}{4\sqrt{3}}[\Gamma_{1}\widehat{\otimes} \Gamma_{1}\widehat{\otimes}\cdots \widehat{\otimes}1_{(2m+1)}+ 1_{(2m+1)}\widehat{\otimes}\Gamma_{1}\widehat{\otimes}\cdots \widehat{\otimes}\Gamma_{1}] \nonumber \\
& & [\gamma_{1}\widehat{\otimes} \gamma_{1}\widehat{\otimes}\cdots \widehat{\otimes}1_{(2m+1)}+ 1_{(2m+1)}\widehat{\otimes}\gamma_{1} \widehat{\otimes}\cdots \widehat{\otimes}\gamma_{1}] \nonumber \\
& & +\frac{1}{4\sqrt{3}}[\Gamma_{2}\widehat{\otimes} \Gamma_{2}\widehat{\otimes}\cdots \widehat{\otimes}1_{(2m+1)}+ 1_{(2m+1)}\widehat{\otimes}\Gamma_{2}\widehat{\otimes}\cdots \widehat{\otimes}\Gamma_{2}] \nonumber \\
& & [\gamma_{2}\widehat{\otimes} \gamma_{2}\widehat{\otimes}\cdots \widehat{\otimes}1_{(2m+1)}+ 1_{(2m+1)}\widehat{\otimes}\gamma_{2}\widehat{\otimes}\cdots \widehat{\otimes}\gamma_{2}] \nonumber \\
& & -\frac{2}{4\sqrt{3}}[\Gamma_{3}\widehat{\otimes} \Gamma_{3}\widehat{\otimes}\cdots \widehat{\otimes}1_{(2m+1)}+ 1_{(2m+1)}\widehat{\otimes}\Gamma_{3}\widehat{\otimes}\cdots \widehat{\otimes}\Gamma_{3}] \nonumber \\
& & [\gamma_{3}\widehat{\otimes} \gamma_{3}\widehat{\otimes}\cdots \widehat{\otimes}1_{(2m+1)}+1_{(2m+1)}\widehat{\otimes}\gamma_{3}\widehat{\otimes}\cdots \widehat{\otimes}\gamma_{3}] \nonumber \\
\end{eqnarray}
The $SU(2)_{L}$ interaction is expressed by:
\begin{eqnarray}
\hspace{-0.7cm}W_{L \mu}^{(n)}&=&\left[\frac{1}{2}(W_{L \mu}^{1} \Gamma_{2} \Gamma_{3}+W_{L \mu}^{2}\Gamma_{1} \Gamma_{3}+W_{L \mu}^{3}\Gamma_{1} \Gamma_{2})\right] \widehat{\otimes}1 \widehat{\otimes} \cdots \widehat{\otimes}1\nonumber \\
\hspace{-0.7cm}& & +\ \cdots \ +1\widehat{\otimes}\cdots \widehat{\otimes}\left[\frac{1}{2}(W_{L \mu}^{1} \Gamma_{2} \Gamma_{3}+W_{L \mu}^{2}+\Gamma_{1} \Gamma_{3}+W_{L \mu}^{3}\Gamma_{1} \Gamma_{2})\right]
\end{eqnarray}
Alternatively,
\begin{eqnarray}
W_{L \mu}^{(n)}&=& \frac{1}{2}W_{L \mu}^{1}(\Gamma_{2}\widehat{\otimes} \Gamma_{2}\widehat{\otimes} \cdots \widehat{\otimes} 1_{(2m+1)}+1_{(2m+1)}\widehat{\otimes}\Gamma_{2}\widehat{\otimes}\cdots \widehat{\otimes}\Gamma_{2}) \nonumber \\
& & (\Gamma_{3}\widehat{\otimes} \Gamma_{3}\widehat{\otimes}\cdots \widehat{\otimes} 1_{(2m+1)}+1_{(2m+1)}\widehat{\otimes}\Gamma_{3}\widehat{\otimes}\cdots \widehat{\otimes}\Gamma_{3}) \nonumber \\
& & +\frac{1}{2}W_{L \mu}^{2}(\Gamma_{1}\widehat{\otimes} \Gamma_{1}\widehat{\otimes}\cdots \widehat{\otimes} 1_{(2m+1)}+1_{(2m+1)}\widehat{\otimes}\Gamma_{1}\widehat{\otimes}\cdots \widehat{\otimes}\Gamma_{1}) \nonumber \\
& & (\Gamma_{3}\widehat{\otimes} \Gamma_{3}\widehat{\otimes}\cdots \widehat{\otimes} 1_{(2m+1)}+1_{(2m+1)}\widehat{\otimes}\Gamma_{3}\cdots \widehat{\otimes}\Gamma_{3}) \nonumber \\
& & +\frac{1}{2}W_{L \mu}^{3}(\Gamma_{1}\widehat{\otimes} \Gamma_{1} \widehat{\otimes}\cdots \widehat{\otimes} 1_{(2m+1)}+1_{(2m+1)}\widehat{\otimes}\Gamma_{1}\widehat{\otimes}\cdots \widehat{\otimes}\Gamma_{1}) \nonumber \\
& & (\Gamma_{2}\widehat{\otimes} \Gamma_{2}\widehat{\otimes}\cdots \widehat{\otimes} 1_{(2m+1)}+1_{(2m+1)}\widehat{\otimes}\Gamma_{2}\widehat{\otimes}\cdots \widehat{\otimes}\Gamma_{2})
\end{eqnarray}
The $U(1)_{R}$ right handed weak interaction is given by
\begin{eqnarray}
W_{R \mu}^{(n)} =\left[\frac{1}{2}(W_{R \mu}^{3} \Gamma_{1} \Gamma_{2})\right] \widehat{\otimes}1\cdots \widehat{\otimes}1+\ \cdots
\ +1\widehat{\otimes}\cdots \widehat{\otimes}\left[\frac{1}{2}(W_{R \mu}^{3} \Gamma_{1} \Gamma_{2})\right]
\end{eqnarray}
or alternatively,
\begin{eqnarray}
 W_{R \mu}^{(n)}&=&\frac{1}{2}W_{R \mu}^{3} (\Gamma_{1}\widehat{\otimes}\Gamma_{1}\widehat{\otimes}\cdots \widehat{\otimes}1_{(2m+1)}+1_{(2m+1)}\widehat{\otimes}\Gamma_{1}\widehat{\otimes}\cdots \widehat{\otimes}\Gamma_{1}) \nonumber \\
& & (\Gamma_{2}\widehat{\otimes}\Gamma_{2}\widehat{\otimes}\cdots \widehat{\otimes}1_{(2m+1)}+1_{(2m+1)}\widehat{\otimes}\Gamma_{2}\widehat{\otimes}\cdots \widehat{\otimes}\Gamma_{2})
\end{eqnarray}
The $U(1)_{B-L}$ interaction $W_{BL \mu}$ can be written as:
\begin{eqnarray}
W_{BL \mu}^{(n)}=\left[\frac{1}{6}(W_{BL \mu}^{ J } \gamma_{1} \Gamma_{1}+W_{BL \mu}^{  J }\gamma_{2} \Gamma_{2}+W_{BL \mu}^{ J }\gamma_{3} \Gamma_{3})\right] \widehat{\otimes}1\cdots \widehat{\otimes}1+\cdots \nonumber \\
+1\widehat{\otimes}\cdots \widehat{\otimes}\left[\frac{1}{6}(W_{BL \mu}^{  J } \gamma_{1} \Gamma_{1}+W_{BL \mu}^{  J }+\gamma_{2} \Gamma_{2}+W_{BL \mu}^{ J }\gamma_{3} \Gamma_{3})\right].
\end{eqnarray}
Alternatively,
\begin{eqnarray}
W_{BL \mu}^{(n)}&=& \frac{1}{6}W_{BL \mu}^{ J }(\gamma_{1}\widehat{\otimes} \gamma_{1}\widehat{\otimes}\cdots \widehat{\otimes} 1_{(2m+1)}+1_{(2m+1)}\widehat{\otimes}\gamma_{1}\widehat{\otimes}\cdots \widehat{\otimes}\gamma_{1}) \nonumber \\
& &  (\Gamma_{1}\widehat{\otimes} \Gamma_{1}\widehat{\otimes}\cdots \widehat{\otimes} 1_{(2m+1)}+1_{(2m+1)}\widehat{\otimes}\Gamma_{1}\widehat{\otimes}\cdots \widehat{\otimes}\Gamma_{1}) \nonumber \\
& & +\frac{1}{6}W_{BL \mu}^{  J }(\gamma_{2}\widehat{\otimes} \gamma_{2}\widehat{\otimes}\cdots \widehat{\otimes} 1_{(2m+1)}+1_{(2m+1)}\widehat{\otimes}\gamma_{2}\widehat{\otimes}\cdots \widehat{\otimes}\gamma_{2}) \nonumber \\
& &  (\Gamma_{2}\widehat{\otimes} \Gamma_{2}\widehat{\otimes}\cdots \widehat{\otimes} 1_{(2m+1)}+1_{(2m+1)}\widehat{\otimes}\Gamma_{2}\widehat{\otimes}\cdots \widehat{\otimes}\Gamma_{2}) \nonumber \\
& & +\frac{1}{6}W_{BL \mu}^{ J }(\gamma_{3}\widehat{\otimes} \gamma_{3}\widehat{\otimes}\cdots \widehat{\otimes} 1_{(2m+1)}+1_{(2m+1)}\widehat{\otimes}\gamma_{3}\widehat{\otimes}\cdots \widehat{\otimes}\gamma_{3}) \nonumber \\
& &  (\Gamma_{3}\widehat{\otimes} \Gamma_{3}\widehat{\otimes}\cdots \widehat{\otimes} 1_{(2m+1)}+1_{(2m+1)}\widehat{\otimes}\Gamma_{3}\widehat{\otimes}\cdots \widehat{\otimes}\Gamma_{3}) \nonumber \\
\end{eqnarray}
Several aspects related to the formulation based on reference \cite{Wei} can be explored. For example, color projection operators, quark projection operators and lepton projection operators can be obtained in our multipartite scenario. However, our goal was to characterize the symmetries associated to the model by exploring two proposals of realization of the elements.

\section{Conclusions} \label{sec:conclu}
In conclusion, we deduce Einstein's equations in a formulation based on geometric algebra using the gauge gravity \cite{Lasenby} of a multipartite perspective. We interpret the gauge fields as fields associated to bosonic and fermionic states from which Einstein's equations emerge. Such fields may be related with qubits and entangled states can be obtained from minimal left ideals using primitive idempotents. Therefore this point of view  also suggests that  gravity may appears as an emerging phenomenon. Using as key ingredients the generalized graded tensor product and the concept of primitive elements extensively explored in the context of Hopf algebras from Lie algebras, an underlying new algebraic structure of the Hopf type is deduced from the formalism so that the gauge fields as well as the qubits can be fully characterized. Applications for background black holes were performed by deducing a multipartite Dirac equation \cite{Lasenby2}. In the context of relativistic quantum information, Bittencourt et al. \cite{Bit1} investigated, through of Dirac bispinors,  when different inertial observers will see different superpositions of quantum states described by Dirac equation solutions (for spin 1/2 states) and how the quantum entanglement changes in this scenario. The Dirac bispinors carry two qubits in a representation supported by $SU(2) \otimes SU(2)$. The negativity and Meyer-Wallach global measures were used for quantify of entanglement. In another recent article \cite{Bit2}, the same authors explore the intrinsic entangled structure associated to Dirac equation in the bilayer graphene. A realistic setup was purposed with the effect of a noise implemented by Ornstein-Uhlenbeck process. In our work, each element of minimal left ideal in $Cl(1,3)$ algebra describes a single qubit since the isomorphism $Cl^{+}(1,3)\simeq Cl(3,0)$ is considered (a representation for $Cl(3,0)$ is given by the Pauli matrices, which correspond to logic gates of a qubit). Two qubits can be described as elements of minimal left ideal in $Cl^{+}(1,3)\otimes Cl^{+}(1,3)\simeq Cl(3,0) \otimes Cl(3,0)$. Particularly, we explore bipartite states from representations in the algebra $Cl^{(2)}(1,3)$ in order to keep relationship of gravity gauge theory. In our formulation, the informational content of Dirac equation is the same as Einstein equations since the Dirac equation is derived through the gauge fields $g^{\mu\nu}$ (this field ensure the equation remain covariant under arbitrary displacement) and $\Omega^{\mu}$ (this field ensure the equation remain covariant under arbitrary displacement) by means of the covariant derivative. The symmetries of the $SU(3)_{c} \times SU(2)_{L} \times U(1)_{R} \times U(1)_{B-R} $ model \cite{Wei} were also explored under two distinct ways in this multipartite formulation. As perspectives we intend to interpret which particles would be associated with the above mentioned fields through the spin. Also, We intend to investigate solutions related to equation (4.3), considering that we believe to be useful in the analysis of Hawking radiation. Another interesting question is the role of quantum entanglement in this formalism.

\section{Acknowledgements}
Eric Pinto thanks to FAPESB and CNPq for partial financial support. We would also like to thank the anonymous referees for their detailed comments and suggestions.


\begin{thebibliography}{}

\bibitem{Waldyr1}{Rodrigues, W.A., Oliveira, E.C.: The Many Faces of Maxwell, Dirac and Einstein
Equations: A Clifford Bundle Approach. Springer, New York (2017)}

\bibitem{Lasenby}{Lasenby, A. N., Doran, C. J. L, and  Gull, S.F.: Gravity, gauge theories and geometric
algebra. Phil. Trans. R. Soc. Lond. A, \href{http://rsta.royalsocietypublishing.org/content/356/1737/487.short}{\textbf{582}, 356:487 (1998).}}

\bibitem{Doran}{Doran., C. J. L.: Geometric Algebra and its Application to Mathematical
Physics. PhD thesis, Cambridge University  (1994).}

\bibitem{strings1}{Zwiebach, B.: A First Course in String Theory. Cambridge University Press, New York (2004)}

\bibitem{loop1}{Thiemann, T.:  Loop Quantum Gravity: An Inside View. Approaches to Fundamental Physics. Lecture Notes in Physics. \href{https://link.springer.com/chapter/10.1007\%2F978-3-540-71117-9_10}{\textbf{721}, 185-263 (2006).}}

\bibitem{Ran1}{Faulkner, T., Guica, M.,  Hartman, T., Myers, R.C. and Van Raamsdonk,M.: Gravitation from Entanglement in Holographic CFTs, JHEP \href{https://link.springer.com/article/10.1007/JHEP03(2014)051}{\textbf{03}, 051 (2014).}}

\bibitem{Ran2}{Lashkari,N.,  McDermott, M. B. and Van Raamsdonk, M.: Gravitational dynamics
from entanglement thermodynamics, JHEP \href{https://link.springer.com/article/10.1007/JHEP04(2014)195}{\textbf{04}, 195 (2014).}}

\bibitem{Ran3}{Lashkari, N., Rabideau, C., Sabella-Garnier, P. and Van Raamsdonk, M.
Inviolable energy conditions from entanglement inequalities, JHEP \href{https://link.springer.com/article/10.1007/JHEP06(2015)067}{\textbf{06}, 067 (2015).}}

\bibitem{Hestenes1}{Hestenes, D. and Sobczyk, G: Clifford Algebra to Geometric Calculus, Reidel, Dordrecht (1984).}

\bibitem{Hestenes2}{Hestenes, D.: Space Time Algebra, Gordon and Breach, New York (1996).}

\bibitem{Verlinde1}{Verlinde, E. P.: On the Origin of Gravity and the Laws of Newton, JHEP \href{https://link.springer.com/article/10.1007/JHEP04(2011)029}{\textbf{1104}, 029 (2011).}}

\bibitem{Verlinde2}{Verlinde, E. P.: Emergent Gravity and the Dark Universe, SciPost Phys. \href{https://scipost.org/SciPostPhys.2.3.016}{\textbf{2}, 016 (2017).}}

\bibitem{Sk}{Sakharov, A. D.: Vacuum Quantum Fluctuations In Curved Space And The Theory
Of Gravitation, Sov. Phys. Dokl. \href{http://iopscience.iop.org/article/10.1070/PU1991v034n05ABEH002498/meta}{\textbf{12}, 1040 (1968).}}

\bibitem{Majid1}{Majid, S.: Hopf algebras for physics at the Planck scale, Classical and Quantum Gravity, \href{http://iopscience.iop.org/article/10.1088/0264-9381/5/12/010/meta}{\textbf{5}, 12, 1587-1607 (1988).}}

\bibitem{Castro}{Castro, C.: Extended Lorentz Transformations in Clifford Space Relativity Theory, Adv. Appl. Clifford Algebras \href{https://link.springer.com/article/10.1007/s00006-015-0529-x}{ \textbf{25} 553-567 (2015).}}

\bibitem{Majid2}{Majid, S.: Foundations of Quantum Group Theory, Cambridge University Press, Cambridge (2000).}

\bibitem{Kassel}{Kassel, C.: Quantum Groups, Springer-Verlag, New York (1995).}

\bibitem{Chari}{Chari, V. and Pressley, A.: A Guide to Quantum Groups, Cambridge University Press, Cambridge (1994).}

\bibitem{Hsu}{Hsu-Wen Chiang, Yoo-Chieh Hu, Pisin Chen: Quantization of space-time baased on a space time interval operator, Physical Review D, \href{https://journals.aps.org/prd/abstract/10.1103/PhysRevD.93.084043}{ \textbf{93}, 084043 (2016).}}

\bibitem{Roldao1}{Da Rocha, R., Bernardini, A. E. and Vaz Jr.,J.:  Int. J. Geom. Methods Mod. Phys. \href{https://www.sciencedirect.com/science/article/pii/S0370269312009380}{\textbf{07}, 821 (2010).}}

\bibitem{Lopez}{Lopez, E.: Quantum Clifford-Hopf Algebras for Even Dimensions, J. Phys. A \href{http://iopscience.iop.org/article/10.1088/0305-4470/27/3/025/meta}{ \textbf{27} 845-854 (1994).}}

\bibitem{Bulacu}{Bulacu, D.: A Clifford algebra is a weak Hopf algebra in a suitable symmetric monoidal category Journal of Algebra \href{https://www.sciencedirect.com/science/article/pii/S0021869311001013}{ \textbf{332}, 244-284 (2011).}}

\bibitem{Fauser1}{Fauser, B., Oziewicz,Z.: Clifford Hopf gebra for two-dimensional space, Miscellanea Algebraicae, \href{Fauser, B., Oziewicz,Z.: Clifford Hopf gebra for two-dimensional space, Miscel-
lanea Algebraicae, 5, 2:31-42, (2001).}{\textbf{5}, 2:31-42, (2001).}}

\bibitem{Fauser2}{Fauser, B.: Quantum Clifford Hopf Algebra for Quantum Field Theory, Adv.Appl.Clifford Algebras \href{https://link.springer.com/article/10.1007/s00006-003-0012-y}{  \textbf{13} 115-125 (2003). } }

\bibitem{Abla}{Ablamowicz, R. and Fauser,B.:  Clifford and Grassmann Hopf algebras via the BIGEBRA package for Maple, Computer Physics Communications \href{https://link.springer.com/article/10.1007/s00006-003-0012-y}{\textbf{170}, 115-130 (2005).}}

\bibitem{Lasenby2}{Lasenby, A.N., Doran, C.J.L.: Geometric Algebra, Dirac Wavefunctions and Black Holes. In: Bergmann P.G., de Sabbata V. (eds) Advances in the Interplay Between Quantum and Gravity Physics. NATO Science Series (Series II: Mathematics, Physics and Chemistry), \textbf{60},  Springer, Dordrecht (2002).}

\bibitem{Baylis}{Baylis, W.E.: Quantum/classical interface: a geometric from the classical side.
In: Byrnes, J. (ed.) Proceedings of the NATO Advanced Study Institute. Dordrecht
Kluwer Academic, Dordrecht (2004).}

\bibitem{Havel}{Havel, T.F., Doran, C.J.: Interaction and entangled in the multiparticle spacetime
algebra. In: Dorst, L., Doran, C.J., Lasenby, J. (eds.) Applications of
Geometric Algebra in Computer Science and Engineering. Birkhauser, Basel(2002).}

\bibitem{Martel} Martel, K. and Poisson, E.: Regular coordinate systems for Schwarzschild and other spherical spacetimes, American Journal of Physics, \href{https://aapt.scitation.org/doi/abs/10.1119/1.1336836}{ \textbf{69}, 476, 2001.}

\bibitem{Wei}{Lu, W.: A Clifford algebra approach to chiral symmetry breaking and fermion mass hierarchies. International Journal of Modern Physics A \href{https://www.worldscientific.com/doi/abs/10.1142/S0217751X17501597}{\textbf{32}, 1750159 (2017).}}

\bibitem{Rodrigues}{Rodrigues Jr., W. A.,  Maiorino, J.E.: A unified theory for construction of arbitrary speeds
$0<v<\infty$ solutions of the relativistic wave equations, Random Oper. and Stoch. Equ. \href{https://www.degruyter.com/view/j/rose.1996.4.issue-4/rose.1996.4.4.355/rose.1996.4.4.355.xml}{ \textbf{4} 355-400 (1996).}}

\bibitem{Bit1} Bittencourt, A. S. V. V., Bernardini, A. E. and Blasone M.: Global Dirac bispinor entanglement under Lorentz boosts, Physical Review A \href{https://journals.aps.org/pra/abstract/10.1103/PhysRevA.97.032106}{ \textbf{97}, 032106 (2018).}

\bibitem{Bit2} Bittencourt, A. S. V. V., Bernardini, A. E. and Blasone M.: Bilayer graphene lattice-layer entanglement in the presence of non-Markovian phase noise, Physical Review B \href{https://journals.aps.org/prb/abstract/10.1103/PhysRevB.97.125435}{ \textbf{97}, 125435 (2018).}


\end{thebibliography}
\end{document}